# Wavefront engineering for controlled structuring of far-field intensity and phase patterns from multimodal optical fibers

Liam Collard,[1,*] Filippo Pisano,[1] Marco Pisanello,[1] Antonio Balena,[1,2] Massimo De Vittorio,[1,2,*] and Ferruccio Pisanello [1,*]

[1] Istituto Italiano di Tecnologia, Center for Biomolecular Nanotechnologies, Arnesano (LE), 73100, Italy

[2] Dipartimento di Ingegneria Dell'Innovazione, Università del Salento, Lecce, 73100, Italy

*Corresponding authors: Liam.Collard@iit.it, Ferruccio.Pisanello@iit.it, and Massimo.DeVittorio@iit.it

**Adaptive optics methods have long been used to perform complex light shaping at the output of a multimode fiber (MMF), with the specific aim of controlling the emitted beam in the near-field. Gaining control of other emission properties, including the far-field pattern and the phase of the generated beam, would open up the possibility for MMFs to act as miniaturized beam splitting, steering components and to implement phase-encoded imaging and sensing. In this study, we employ phase modulation at the input of a MMF to generate multiple, low divergence rays with controlled angles and phase, showing how wavefront engineering can enable beam steering and phase-encoded applications through MMFs.**

Multimode optical fibers (MMFs) are intrinsically turbid media, and the transmission of light through them is well known to scramble the phase and amplitude distribution of coherent light entering in the proximal facet of the waveguide. As such, there has been significant interest in developing methods to deterministically pre-shape the phase of coherent light radiation prior to transmission through the fiber to control the emission patterns at the distal end. This approach recently allowed the realization of low-invasiveness endoscopes based by a single imaging MMF. Several adaptive optics methods have been applied to perform high-resolution imaging through wavefront modulation, including spatial light modulator (SLM) sub-domain optimisation, SLM-based plane wave generation [1–3], digital micromirror device (DMD) phase optimization [4] and digital phase conjugation [5]. All these methods enable structuring light emission in the vicinity of the fiber facet, e.g. in the near field. This allowed to obtain a series of high-resolution imaging methods in transmission, reflection fluorescence modes [6] , Raman imaging [7] and Coherent Anti-Stokes Raman Spectroscopy (CARS) [8,9], optical manipulation [10] and through multicore fibers [11].

Together with near-field structuring, far-field engineering of coherent light emitted from MMF also has peculiar potentials. As transmitted amplitude is scrambled, the same happens to output angles, resulting in speckle patterns also in the far-field emission diagram. Controlling the light field at infinity offers interesting perspectives as the fiber output can be collimated at arbitrary angles within the fiber NA. This, in turn, opens the way to the exploitation of multimode fibers as passive beam-steering elements and beam splitting components. Notably, this can be achieved with a standard flat-cleaved fiber and does not require fabricating a meta-surface on the fiber facet [12,13]. Additionally, as any variation in the near-field distribution due to external events is reflected in the far-field structure, studying this mechanism shows potential in the development of remote sensing techniques.

In this study, we apply a plane wave optimisation with spatial light modulation to control the far-field emission pattern from a MMF. The system is employed to tailor the output angle from the waveguide with its NA and low divergence, to generate multiple beams emerging from the facet at different angles and to tailor their phase.

The implemented optical setup to control far-field emission patterns from a MMF is displayed in Figure 1A. A continuous wave 632 nm laser beam was expanded to overfill the 512 × 512 pixel screen of a SLM (ODP512, Meadowlark optics). The SLM displays a phase pattern that generates a scanning beam and a reference beam

(hereafter referred to as $B_{scan}$ and $B_{ref}$), both sent to lens L1 (focal length $f_{L1}$=200 mm). A beam splitter (BS1) was placed 10 cm from L1 to split the optical path in a *scan path* (transmitted by BS1), which exploits $B_{scan}$, and a *reference path* (reflected by BS1), instead exploiting $B_{ref}$.

On the *scan path*, lenses L1 and L2 ($f_{L2}$=200 mm) form a *4f* relay between the SLM and the back aperture of the microscope objective MO1 (0.65 NA, 40x, AMEP4625, ThermoFisher), while BS2 is used to image the input facet of the MMF onto a camera (CCD1). MO1 focuses the beam onto a MMF (NA 0.22, core diameter 50 μm, Thorlabs FG050UGA), which randomizes output pattern resulting in new beam referred to as $B_{random}$. Light transmitted through the MMF was collected by the microscope objective MO2 (10x, 0.3 NA, MPLFLN10x – Olympus, $f_{MO2}$=18 mm) and lenses L4 ($f_{L4}$=125 mm) and L5 ($f_{L5}$=100 mm) were positioned such that L4 and L5 form a 4f optical system conjugating the back aperture of MO2 on CCD2 (CS505MU - Thorlabs), projecting on it the far-field of the fiber facet (referred to as *(u,v)* plane). Each point of the *(u,v)* plane represents an output angle from the MMF, following the relation:

$$r = \frac{f_5}{f_4} f_{MO2} \tan \theta_{out} \quad (1)$$

where $\theta_{out}$ is the angle of the beam emitted by the fiber measured with respect to the optical axis, $r = \sqrt{u^2 + v^2}$ is the distance of the generic point *(u,v)* from the center of the far-field. Or in component form, along the *u* and *v* axis: $u = \frac{f_5}{f_4} f_{MO2} \tan \theta_u$ and $v = \frac{f_5}{f_4} f_{MO2} \tan \theta_v$. On the *reference path*, an adjustable iris was placed at the focal point of L1 blocking the residual scanning beam. $B_{ref}$ was then expanded by a beam expander formed by lenses L7 and L8 ($f_{L7}$=30 mm, $f_{L8}$=100 mm), before being re-joined with the scanning beam by BS3. Thus, $B_{random}$ and $B_{ref}$ beams interfere on CCD2 with an angle controlled by BS3 (see representative interferogram in Figure 1b).

To obtain the $B_{scan}$ and $B_{ref}$ beams, blazed sawtooth gratings $\phi_{scan}$ and $\phi_{ref}$ were generated and summed together to obtain the image $\phi_{interfere}^{j_x,j_y,p}(x,y)$ on the SLM, whose screen plane is referred by *x* and *y* axes, with:

$$\phi_{scan}^{j_x,j_y,p}(x,y) = mod\,(ax + by + p, 2\pi),$$

$$\phi_{ref}(x,y) = mod(cx, 2\pi)$$

Where, *a* and *b* define the periodicity of the grating along *x* and *y* so that the beam $B_{scan}$ can scan on a 25 x 25 grid in the input facet of the MMF where each element of this grid is identified by the indexes $(j_x, j_y)$ (see bottom left inset in Figure 1a). *c* defines the periodicity of $\phi_{ref}$ and it was chosen so that $B_{ref}$ is aligned with the iris on the *reference path* and not on the core of the MMF along the *scan path*. *p* is the phase shift applied at each position in the $(j_x, j_y)$ plane in the procedure described below.

To obtain a specific emission angle $\theta_{out}$ from the MMF, a calibration procedure was applied, based on the method proposed in [1]. In the algorithm the $(u,v)|_{\theta_{out}}$ point related to a specific $\theta_{out}$ is monitored in intensity on the interference pattern on CCD2 while scanning the beam on the $(j_x, j_y)$ plane using the following image as an input to the SLM

$$\phi_{interfere}^{j_x,j_y,p}(x,y) = \arg(\exp\left(i\phi_{scan}^{j_x,j_y,p}(x,y)\right) + \exp(i\phi_{ref})).$$

For each $(j_x, j_y)$ pair, *p* was tested from 0 to $2\pi$ (the full scan procedure in the input facet is shown in visualization 1). The intensity variation as a function of *p* on a representative $(u,v)|_{\theta_{out}}$ point of the far-field plane is displayed in Figure 1C, from which the phase step $p_{opt}^{j_x,j_y}$ giving the maximum intensity value at the coordinate *(u,v)* is extracted and used to generate the phase pattern $\phi_{opt}^{j_x,j_y}(x,y) = \phi_{scan}^{j_x,j_y,p_{opt}^{j_x,j_y}}(x,y)$ for the $j_x$-th, $j_y$-th grid element.

This input-phase optimization procedure is repeated for all the 625 points on the $(j_x, j_y)$ plane, generating the SLM input image to focus light in the point *(u,v)* of the far field, as given by the sum of all optimized phase patterns

$$\Phi(x,y) = \arg\left(\sum_{j_x=1}^{25}\sum_{j_y=1}^{25} \exp\left(i\phi_{opt}^{j_x,j_y}(x,y)\right)\right). \quad (2)$$

A typical result of this optimization is shown in Figure 1D for different *(u,v)* far field points, while a scanning spot over the far field plane is shown in visualization 2. Concerning computational times, if a 30 by 30 square array of pixels is targeted on the *(u,v)* plane and calculations are run on a graphic processing unit (GPU, GeForce GTX 960 – NVIDIA in our case), the entire process was completed within approximately 15 minutes. On average, about 9% of the total intensity transmitted through the MMF is contained within the focused spot in the far field. To image the near-field lens L4 was removed, the intensity distribution appears to be stochastic and no trace of the far-field focus (shown in 1E) is visible (see supplementary figure S1).

To measure the azimuthal angle $\theta_{out}$ and divergence angle $\alpha$ of the beam emitted by the fiber, MO2 was removed and a CCD (CS505MU – Thorlabs) fixed to a single axis translation stage was positioned approximately 5 mm from the output of the MMF as shown in Figure 2A. A set of images of the 30 by 30 array of far-field spots were recorded, the CCD was then moved 1 mm closer to the fiber and another set of the same size was acquired. Error limits were calculated by considering the distance between the two positions to be 1.0±0.1mm to account for error in the position of CCD2. To calculate $\theta_{out}$ we considered the coordinate difference of the maximum intensity pixel for the first set $(u_1, v_1)$ and second set $(u_2, v_2)$ of images. Two example images in each position are shown in the inset of figure 2A. The pixel size on the CCD was 3.45 µm by 3.45 µm, therefore:

$$\theta_{out} = \tan^{-1}\left(\frac{3.45\sqrt{(u_1-u_2)^2 + (v_1-v_2)^2}}{1000}\right) \quad (4)$$

To calculate the divergence angle $\alpha$ consider the full width at half maximum of both spots, $FWHM_1$ and $FWHM_2$, then

$$\alpha = 2\tan^{-1}\left(\frac{3.45(FWHM_1 - FWHM_2)}{1000}\right) \quad (5)$$

Figure 2B is a color-map showing the measured $\theta_{out}$ for the targeted point of the *(u,v)* plane. As expected, at the centre of the far-field the azimuthal angle is close to zero, whereas at the edge $\theta_{out}$ approaches the limit of the numerical aperture of the fiber (corresponding to 12.7°). The relationship between the far field coordinate and azimuthal angle is expected to follow equation 1. The experimental relationship between the far field coordinate and the measured $\theta_{out}$ is shown in the profile plot in figure 2C. The angular components in the *u* and *v* direction are shown in Figure 2D and E calculated as:

$$\theta_{out_u} = \tan^{-1}\left(\frac{3.45(u_1 - u_2)}{1000}\right)$$

$$\theta_{out_v} = \tan^{-1}\left(\frac{3.45(v_1 - v_2)}{1000}\right)$$

The divergence angle $\alpha$ is plotted as a colour-map in figure 2F. A histogram of the data in Figure 2F is shown in Panel G, featuring over 80% of the points between -1° and 1° (average 0.70° standard deviation 0.68°).

After demonstrating how far-field beam steering of a laser beam can be performed through a MMF, next we show how beam splitting may be performed to simultaneously generate *n* multiple beams $f_{\theta_k}$ exiting the fibre with azimuthal angle $\theta_j$. If $p_{opt}^{j_x,j_y,k}$ is the optimised phase shift for $k^{th}$ far-field focus, with $k$ ranging from 1 to n, the phase image on the SLM is then:

$$\Phi = \arg\left(\sum_{j_x=1}^{25}\sum_{j_y=1}^{25}\sum_{k=1}^{n} \exp(i\phi_{scan}^{j_x,j_y,p_{opt}^{j_x,j_y,k}})\right)$$

Figure 3 shows the far-field image of three multiple beam arrays with 1, 2 and 3 different angles simultaneously. In this example between 4% and 6% of the total intensity is contained in each hot-spot array, with a signal to noise ratio reduced as the number of beams is increased: the average beam intensity ($\frac{1}{n}\sum_{k=1}^{n} B_{\theta_k}$) falls from 1 to 0.5 to 0.39 and the signal to noise ratio falls from 268 to 132 to 108. Therefore, a far-field optimisation of light transmission through a MMF can be used to perform complex beam steering of multiple beams, with then above-detailed limitations in terms of divergence angle and signal to noise ratio.

Another parameter that can be controlled by the optimisation procedure is the phase of the outputted beam. So far, we have optimised the phase of the input beam to $p_{opt}^{j_x,j_y}$ which gives the maximum intensity of the interference fringes at the targeted spot *(u,v)*, i.e giving constructive interference between $B_{ref}$ and $B_{out}$. However, it is also possible for $B_{ref}$ and $B_{out}$ to be out of phase by *p* where *p* is in the range 0-2π. This can be done by applying a shift *p* to $p_{opt}^{j_x,j_y}$ i.e. the phase is shifted by $q = mod(p_{opt}^{j_x,j_y} + p, 2\pi)$. This gives an input phase image

$$\Phi_{phase}(x,y,p) = \arg\left(\sum_{j_x=1}^{25}\sum_{j_y=1}^{25} \exp\left(i\phi_{scan}^{j_x,j_y,q}(x,y)\right)\right),$$

generating beam $B_{phase}(p)$. 16 phase steps were applied to a single focused spot in the far-field plane. The intensity distribution is unaffected by the application of phase shift *p*. To measure the induced phase shift, the phase $\psi$ of $B_{phase}(p)$ was reconstructed from the Fourier transform of the interferogram between $B_{phase}(p)$ and $B_{ref}$ [5]. The interferogram and the reconstructed phase of $B_{phase}$ for a representative value of p= π are shown in Figures 4A and B, respectively. Figure 4C shows the calculated phase shift $\Delta\psi(p) = \psi\left(B_{phase}(p)\right) - \psi(B_{phase}(0))$. The linear fit of $\Delta\psi(p)$ was $\Delta\psi(p)=0.98p + 0.18$, indicating the phase shift applied by the SLM was well preserved at the modulated beam emitted by the MMF. We have observed that this phase modulation takes place within the entire fiber NA, although the system provides a focused spot in a specific point of the far-field, where the phase values are less dispersed than in the rest of the core (Figure 4D).

In summary, we report evidence that wavefront engineering techniques can be employed to achieve full control on the distribution of intensity and phase of light in the far field of a MMF. Our results can potentially enable pure fiber-based beam steering of multiple collimated beam-lets, and open the door to the development of phase-encoded imaging and sensing through a MMF.

**Funding.** L.C., M.D.V. and F. Pisanello acknowledge funding from the European Union's Horizon 2020 research and innovation program under grant agreement No 828972. F. Pisano, A.B. and F. Pisanello acknowledge funding from the European Research Council under the European Union's Horizon 2020 research and innovation program (G.A.677683). M.P. and M.D.V. acknowledge funding from the European Research Council under the European Union's Horizon 2020 research and innovation program (G.A. 692943). M.P., F. Pisanello, and M.D.V. are funded by the US National Institutes of Health (1UF1NS108177-01). M.D.V. is funded by the US National Institutes of Health (U01NS094190).

See supplement 1 for supporting content.

**References**

1. T. Čižmár and K. Dholakia, "Shaping the light transmission through a multimode optical fibre: complex transformation analysis and applications in biophotonics," Opt. Express **19**, 18871–18884 (2011).

2. M. Plöschner, T. Tyc, and T. Čižmár, "Seeing through chaos in multimode fibres," Nat. Photonics **9**, 529–535 (2015).

3. M. Plöschner, B. Straka, K. Dholakia, and T. Čižmár, "GPU accelerated toolbox for real-time beam-shaping in multimode fibres," Opt. Express **22**, 2933–2947 (2014).

4. A. M. Caravaca-Aguirre, E. Niv, D. B. Conkey, and R. Piestun, "Real-time resilient focusing through a bending multimode fiber," Opt. Express **21**, 12881–12887 (2013).

5. I. N. Papadopoulos, S. Farahi, C. Moser, and D. Psaltis, "Focusing and scanning light through a multimode optical fiber using digital phase conjugation," Opt. Express **20**, 10583–10590 (2012).

6. S. A. Vasquez-Lopez, R. Turcotte, V. Koren, M. Plöschner, Z. Padamsey, M. J. Booth, T. Čižmár, and N. J. Emptage, "Subcellular spatial resolution achieved for deep-brain imaging in vivo using a minimally invasive multimode fiber," Light Sci. Appl. **7**, 110 (2018).

7. I. Gusachenko, M. Chen, and K. Dholakia, "Raman imaging through a single multimode fibre," Opt. Express **25**, 13782–13798 (2017).

8. A. Lombardini, V. Mytskaniuk, S. Sivankutty, E. R. Andresen, X. Chen, J. Wenger, M. Fabert, N. Joly, F. Louradour, A. Kudlinski, and H. Rigneault, "High-resolution multimodal flexible coherent Raman endoscope," Light Sci. Appl. **7**, 10 (2018).

9. J. Trägårdh, T. Pikálek, M. Šerý, T. Meyer, J. Popp, and T. Čižmár, "Label-free CARS microscopy through a multimode fiber endoscope," Opt. Express **27**, 30055–30066 (2019).

10. S. Bianchi and R. Di Leonardo, "A multi-mode fiber probe for holographic micromanipulation and microscopy," Lab Chip **12**, 635–639 (2012).

11. V. Tsvirkun, S. Sivankutty, K. Baudelle, R. Habert, G. Bouwmans, O. Vanvincq, E. R. Andresen, and H. Rigneault, "Flexible lensless endoscope with a conformationally invariant multi-core fiber," Optica **6**, 1185–1189 (2019).

12. A. Xomalis, I. Demirtzioglou, E. Plum, Y. Jung, V. Nalla, C. Lacava, K. F. MacDonald, P. Petropoulos, D. J. Richardson, and N. I. Zheludev, "Fibre-optic metadevice for all-optical signal modulation based on coherent absorption," Nat. Commun. **9**, 182 (2018).

13. N. Yu and F. Capasso, "Optical Metasurfaces and Prospect of Their Applications Including Fiber Optics," J. Light. Technol. **33**, 2344–2358 (2015).

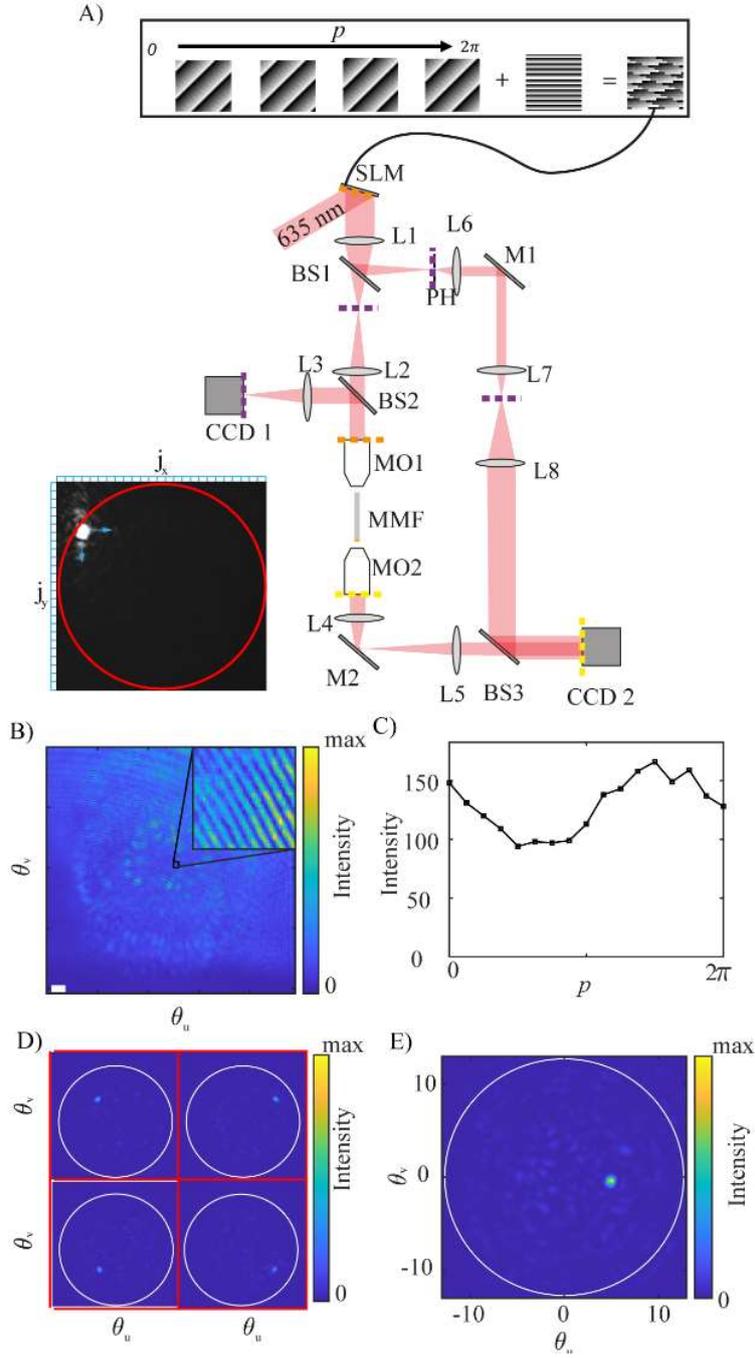

Fig. 1. A) Optical system to generate hot-spots at the output from the MMF, M mirror, L lens, MO objective, MMF multimode fiber, SLM spatial light modulator, BS beam splitter, CCD charged coupling device. Dashed lines show conjugate planes. Top inset: construction of the phase pattern displayed by the SLM to obtain $B_{scan}$ and $B_{ref}$. Bottom left inset: indexing of the input facet of the MMF in a grid of 25x25 pixels elements. B) Example interferogram $B_{interfere}^{j_x,j_y,p}$ C) Intensity variation of $B_{interfere}^{j_x,j_y,p}$ at pixel $(u,v)$ over the phase shift. D) Image on CCD2 of a far-field hot-spot (white circle indicates theoretical limit of the fiber NA). E) Images on CCD2 of multiple far-field hot-spots.

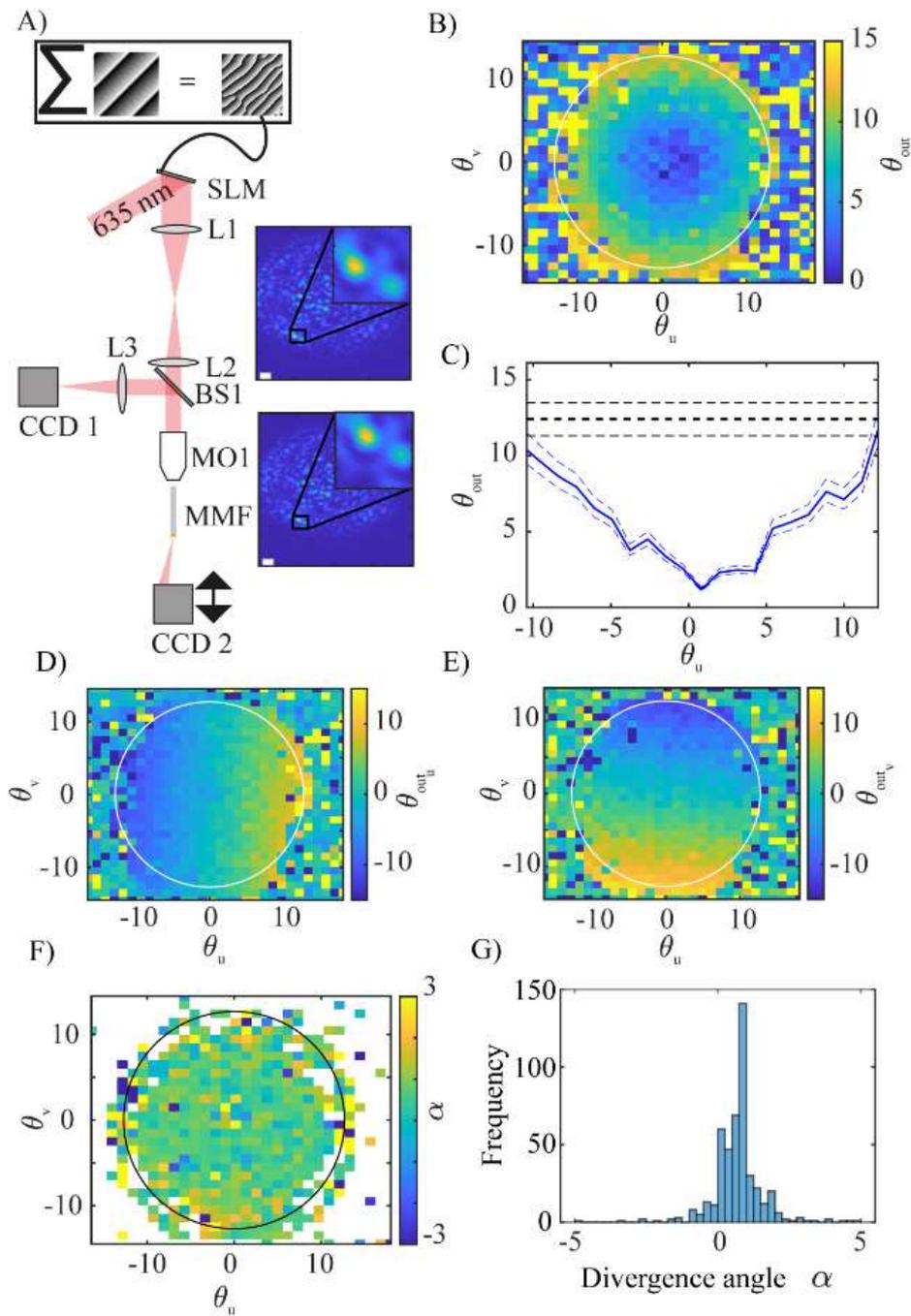

Fig. 2. A) Optical setup to measure $\theta_{out}$ and α CCD2 is attached to a translation stage. B) Measurement of $\theta_{out}$ for 30 by 30 far field hot-spots. C) Line plot of B, blue dashed lines are the error on $\theta_{out}$. Black dashed lines indicate the theoretical limit from the fiber NA . D) Measurement of $\theta_{out_u}$ for 30 by 30 far field hot-spots. E) Measurement of $\theta_{out_v}$ for 30 by 30 far field hot-spots. F) Measurement of divergence angle α for 30x30 far-field hot-spots. Here, points outside of the NA of the core of the MMF have been set to white (by setting a threshold on the intensity ratio of the far field focus). G) Histogram of α.

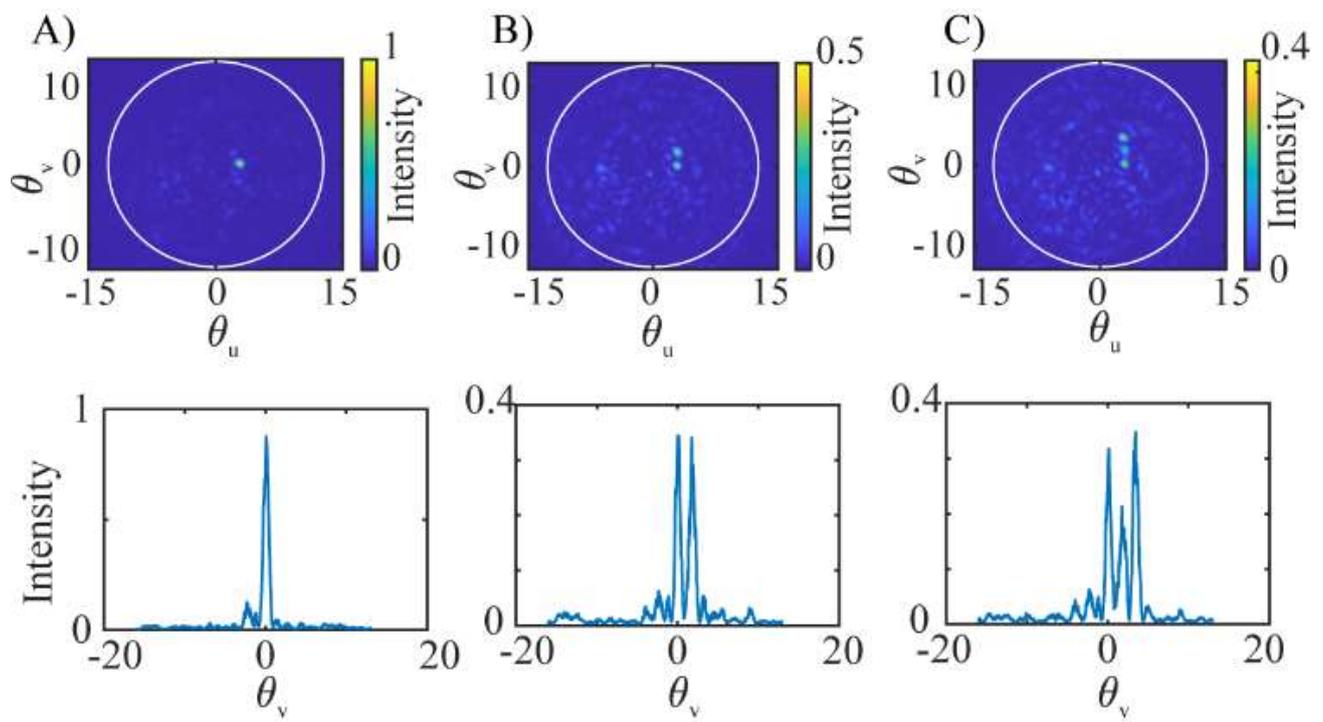

Fig 3. Multiple far-field hot spots generated simultaneously and normalised profile plots (averaged over 20 pixels).

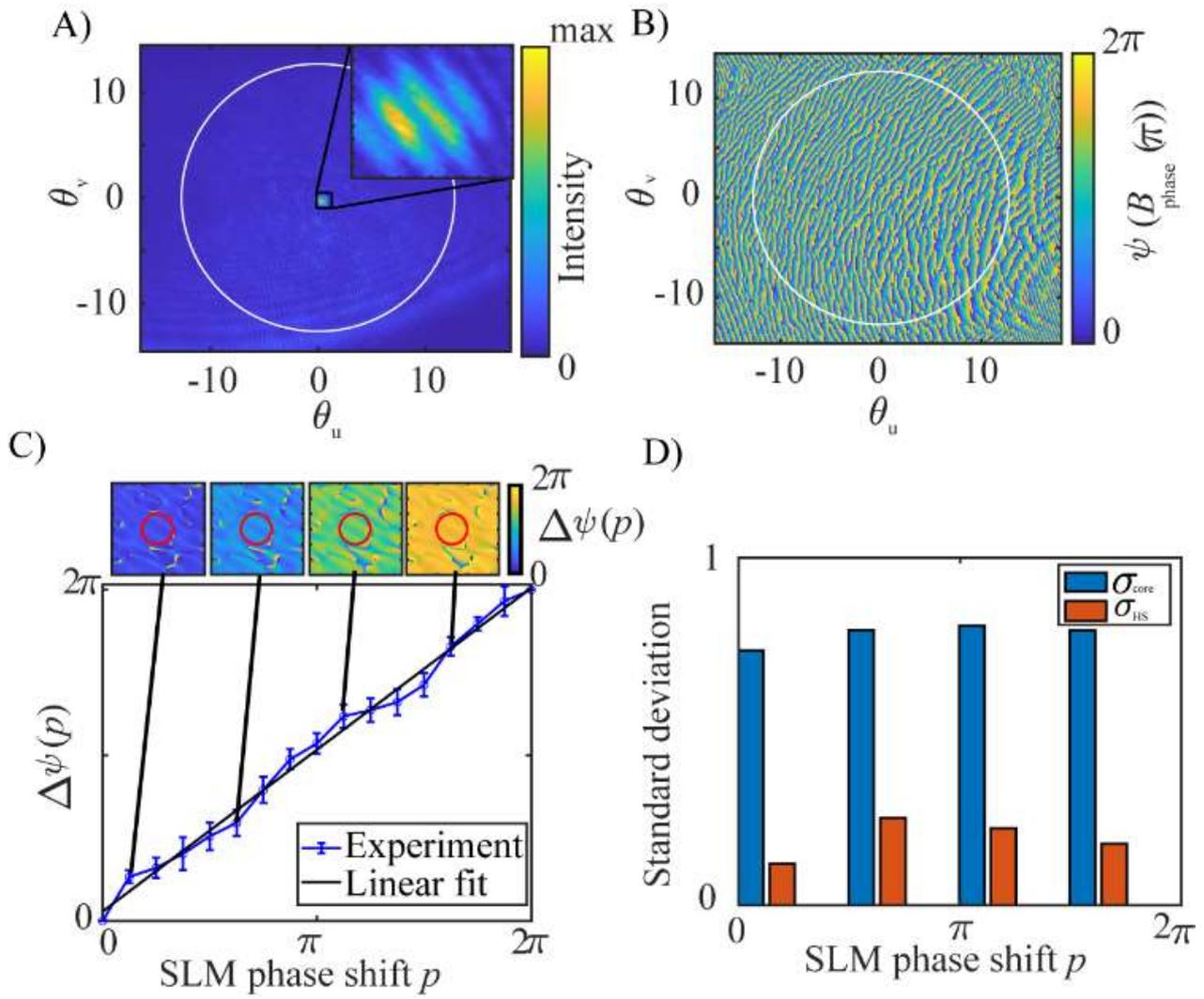

Fig. 4. A) Interferogram between $B_{ref}$ and $B_{phase}$ B) Direct phase measurement of $B_{phas}$ ($\pi$), circle indicates the limit of the NA of the fibre. C) Top row: Measurement of phase shift between $B_{phase}(0)$ and $B_{phase}(p)$. $\Delta\psi(p) = \psi\left(B_{phase}(p)\right) - \psi(B_{phas}(0))$ blue line is the experimental data and black line is a linear fit. D) Bar graph showing the standard deviation of $\Delta\psi(p)$ over the entire core (blue) and in the hotspot region (orange).

# Supplementary information.

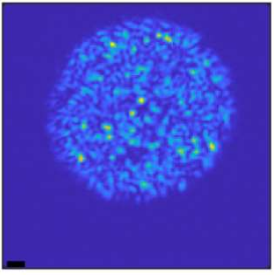

Fig S1 - The near-field image of figure 1E obtained by removing L4 (scale bar 5 μm)